\documentclass[jpcm]{iopart}
\usepackage{graphicx,citesort}

\newcounter{fnt}

\begin{document}

\title{On the shape of barchan dunes}

\author{Klaus Kroy, Sebastian Fischer and Benedikt Obermayer}
\address{Hahn-Meitner Institut, Glienicker Stra\ss e 100, 14109
Berlin, Germany}

\begin{abstract}
Barchans are crescent-shaped sand dunes forming in aride regions with
unidirectional wind and limited sand supply. We report analytical and
numerical results for dune shapes under different environmental
conditions as obtained from the so-called `minimal model' of aeolian
sand dunes. The profiles of longitudinal vertical slices (i.e.\ along
the wind direction) are analyzed as a function of wind speed and sand
supply. Shape transitions can be induced by changes of mass, wind
speed and sand supply.  Within a minimal extension of the model to the
transverse direction the scale-invariant profile of transverse
vertical cuts can be derived analytically.
\end{abstract}


\ead{kroy@hmi.de}

\section{Introduction}
\begin{figure}
   \includegraphics[width=0.5\columnwidth]{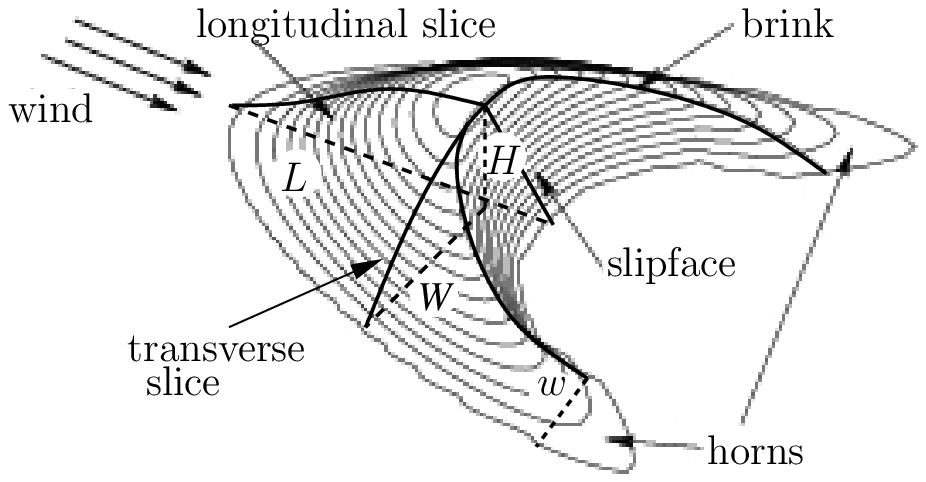}
  \hspace{-0.5\columnwidth} a) \hfill
  b) $\!\!\!\!\!\!\!\!$
  \includegraphics[width=0.5\columnwidth]{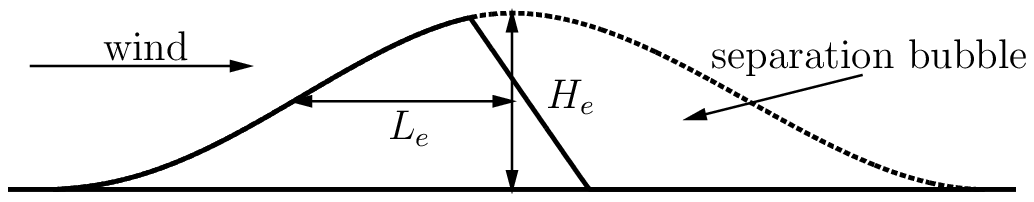}
  \caption{\label{fig:sketch} a)~Sketch of a barchan dune introducing
    some terminology. b)~Longitudinal cut along the symmetry
    plane. The region of flow separation is modelled by a smooth
    continuation of the dune profile
    \cite{kroy-sauermann-herrmann:2002} and the shear stress on the
    back of the dune is calculated as for a heap with the shape of the
    \emph{envelope} (dune and separation bubble) with height $H_{\rm
    e}$ and characteristic length $L_{\rm e}$.}
\end{figure}

The interaction of turbulent wind with sand is known to create a
surprisingly rich variety of sand structures in many places on Earth
and also on other planets. Among the most symmetric structures are
isolated heaps and barchan dunes. For a schematic sketch that
introduces some of the terminology see figure~\ref{fig:sketch}.
By identifying and isolating the mechanisms that are essential for
dune formation, a mathematical minimal model could recently be
formulated
\cite{kroy-sauermann-herrmann:2002,kroy-sauermann-herrmann:2002b}, and
was shown to reproduce generically field observations reporting
systematic shape variations \cite{sauermann-etal:2000}. In particular,
shape transitions from dunes (with slipface) to smooth heaps as a
function of sand mass were predicted
\cite{kroy-sauermann-herrmann:2002,kroy-sauermann-herrmann:2002b,andreotti-claudin-douady:2002}.
The model adapts, combines, and extends earlier theoretical
developments in turbulence
\cite{hunt-leibovich-richards:88,weng-etal:91} and aeolian sand
transport \cite{Owen64,sauermann-kroy-herrmann:2001} into a consistent
mathematical description, which allows to identify a weak spontaneous
symmetry breaking of the turbulent air flow as the origin of the
growth instability giving rise to structure formation. It moreover
pinpoints the mutual competition of this symmetry breaking with
saturation transients in the sand flux as the basic mechanism
responsible for shape selection as a function of size and
environmental parameters such as density of the air, wind speed or
sand supply. The characteristic length scale provided by the
saturation transients is called the \emph{saturation length}
$\ell_{\rm s}$
\cite{sauermann-kroy-herrmann:2001,kroy-sauermann-herrmann:2002,%
kroy-sauermann-herrmann:2002b}. Due to the ballistic grain motion, it
can be much larger than the grain diameter $d$ ($\approx 10^{-4}$ m),
which is the only elementary scale in the problem, and can thus
compete with the turbulent symmetry breaking to cause scale dependence
and shape transitions of dunes that are of the order of $10^2$ m long.

Below, we address some consequences of (broken) scale invariance by
analyzing the predicitions of the minimal model for the longitudinal
shape of a barchan as a function of environmental parameters. The next
section reviews the major parameter dependencies of $\ell_{\rm s}$ and
deduces some immediate consequences for shape
similarities. Section~\ref{sec:long} reports a systematic numerical
study of the influence of dune size, wind speed and sand supply on the
longitudinal shape.  Finally, in section~\ref{sec:trans}, we extend
the minimal model to the transverse direction.  We close with a brief
conclusion.

\section{Similarity hypothesis}\label{sec:sim}
With only one characteristic length $\ell_{\rm s}$ in the problem, one
expects that shapes of dunes of various size and under various
environmental conditions can be related to each other by similarity
transformations.  This immediate consequence of the minimal model
\cite{kroy-sauermann-herrmann:2002,kroy-sauermann-herrmann:2002b} is
supported by careful field measurements \cite{sauermann-etal:2000}
and, more recently, by laboratory experiments with submarine dunes
\cite{hersen-douady-andreotti:2002,kroy-guo:2004,%
endo-kubo-sunamura:2004,endo-taniguchi-katsuki:2004}.
If this promising evidence could be further validated in future work,
submarine structures forming in unidrectional flows, which were
formerly identified as \emph{ripples}
\cite{betat-frette-rehberg:99,betat-etal:2002} on the basis of their
superficial similarity to aeolian ripples, would actually have to be
regarded as proper dunes. This could tremendously facilitate the
experimental study of dunes and ultimately allow us to draw
conlcusions about kilometer-sized dunes on Earth and possibly on other
planets from analyzing centimeter-sized submarine dunes in the lab.
However, some caution is needed in postulating simple one-to-one
relations of similar structures under completely different
environmental conditions. Arguably, the minimal model, which has the
form of a concise hydrodynamic description of sand transport in
turbulent shear flow \cite{kroy-sauermann-herrmann:2002b}, could well
be applicable outside the realm of validity of the expressions and
values derived for its kinetic coefficients, such as $\ell_{\rm s}$,
so that similar dune structures may not always be reliable indicators
for similar elementary transport processes.

For any fixed given ratio $\tau/\tau_{\rm t}$ of the wind shear stress
$\tau$ to the threshold shear stress $\tau_{\rm t}$ for the
mobilization of grains by the wind, the saturation length is predicted
\cite{sauermann-kroy-herrmann:2001,kroy-sauermann-herrmann:2002b} to
scale linearly in $\tau_{\rm t}$ divided by the gravitational
acceleration $g$ and the atmospheric density $\rho_{\rm a}$. According
to conventional estimates from simple arguments for the aerodynamics
on the grain scale, $\tau_{\rm t}=\theta\rho' g d$ \cite{Bagnold41},
i.e.\ $\ell_{\rm s}$ scales linearly in the grain diameter $d$ as
required by dimensional analysis and also linearly in $\rho'/\rho_{\rm
a}$, the immersed over the atmospheric density. The dimensionless
(generalised) Shields parameter $\theta$ \cite{shields:1936,Bagnold41}
accounts for the complicated interactions of the hopping grains with
the air and the bed. It usually can be taken to be a phenomenological
constant for aeolian sand transport on Earth. The density ratio
$\rho'/\rho_{\rm a}$ ranges from about $2\cdot 10^5$ for typical
conditions on Mars, through $2.2\cdot 10^3$ on Earth and about $50$ on
Venus, to $1.7$ for submarine flows and can thus dramatically amplify
the effect of broken scale invariance beyond the grain scale $d$.
Typical grain sizes found in dunes are estimated to be roughly the
same on Mars and Earth \cite{edgett-christensen:91}, and similar (or
slightly smaller) grain sizes have been employed in laboratory
experiments with submarine dunes
\cite{betat-frette-rehberg:99,betat-etal:2002,%
hersen-douady-andreotti:2002}. One can therefore roughly estimate the
ratios of the various $\tau_{\rm t}$ as Mars/Earth/Venus/water
$\approx 10/1/1/0.5$, respectively, if potential differences in the
Shields parameter are disregarded \cite{iversen-white:82}.
Accordingly, we expect $\ell_{\rm s}$ ratios (and thus size ratios, if
size is measured relative to the corresponding grain size) of
\begin{equation}\label{eq:sim}
\mbox{Mars/Earth/Venus/Water} \approx 10^5/10^3/10^1/1
\end{equation}
for dunes of the same shape, with the same $\tau/\tau_{\rm t}$, and
with the same degree of saturation of the upwind sand flux. This
implies that asymptotically large dunes, which should have a universal
scale-invariant shape, are most easily produced in water
\cite{betat-frette-rehberg:99,betat-etal:2002,hersen-douady-andreotti:2002}.
The similarity rule (\ref{eq:sim}) is well supported by a comparison
of aeolian and submarine dunes
\cite{sauermann-etal:2000,hersen-douady-andreotti:2002,kroy-guo:2004}.
However, there may be some evidence \cite{breed:77,Thomas99} that it
overestimates the scale factor between dunes on Earth and on
Mars. This could be a consequence of the considerable differences in
atmospheric densities and corresponding differences in the dominant
sand transport mechanisms and Shields parameters on both planets
\cite{marshall-borucki-bratton:98}.

\begin{figure}
a)$\!\!\!\!\!\!\!\!$\includegraphics[width=0.5\columnwidth]{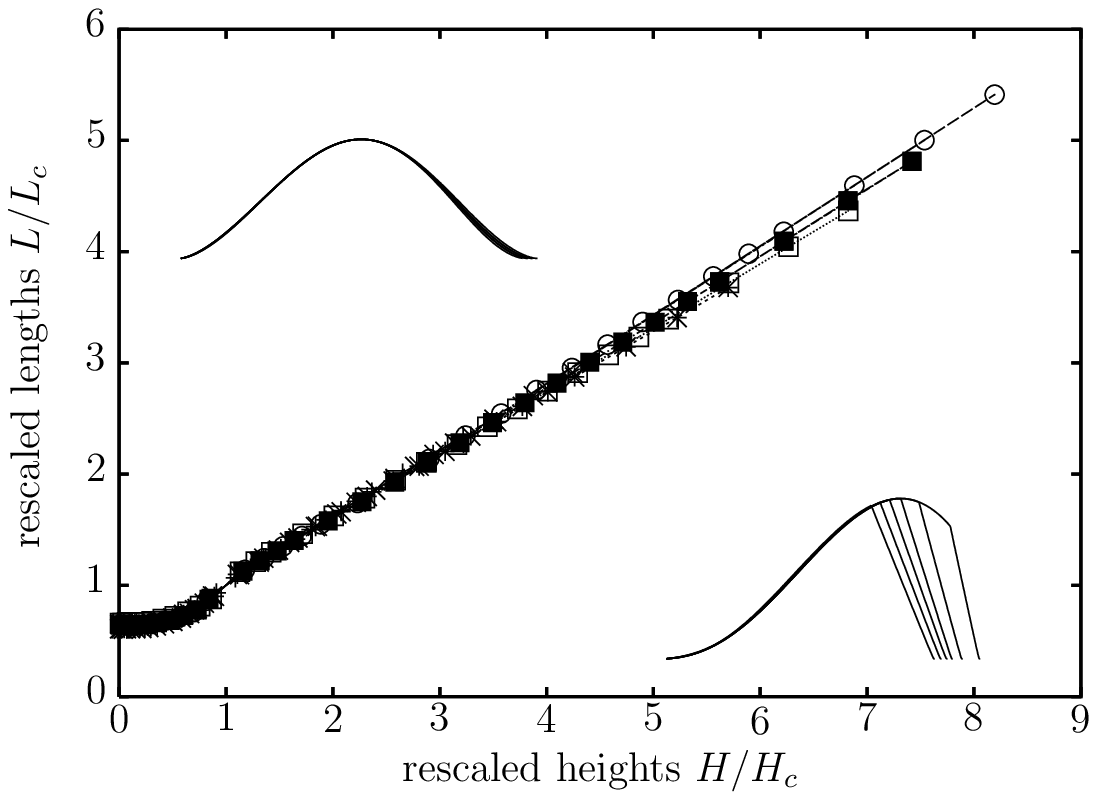}
\quad
b)$\!\!\!\!\!\!\!\!$\includegraphics[width=0.5\columnwidth]{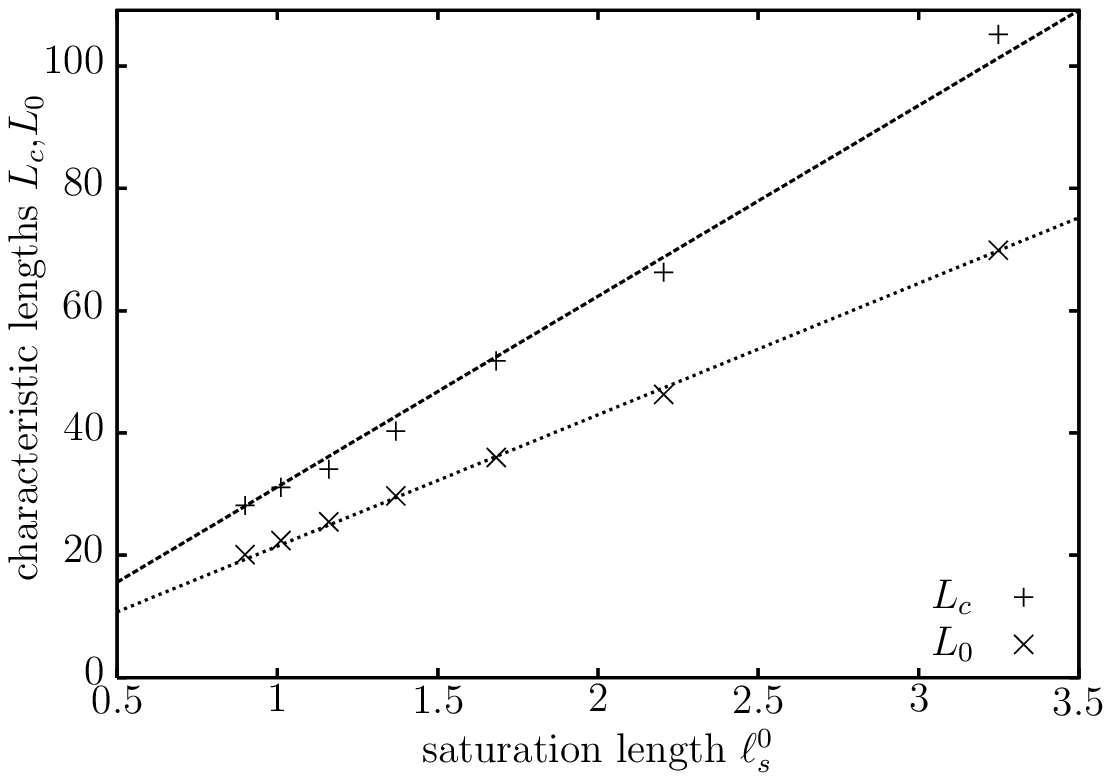}
\caption{\label{fig:steady} a) Height-length relations of windward
  longitudinal steady-state profiles for various ``masses'' ($1-676$
  m$^2$) and wind speeds ($\tau^0/\tau_{\rm t}=1.4$, 1.6, 1.8, 2.0,
  2.2, 2.4, 2.6) collapse onto a universal curve if heights and
  lengths are normalized to their values $H_{\rm c}$, $L_{\rm c}$ at
  the shape transition. \emph{Insets:} The windward profiles of all
  dunes and heaps are of the functional form~(\ref{eq:hx}). b) The
  transition length $L_{\rm c}$ and the minimum heap length
  $L_{0}\approx 0.7L_{\rm c}$ scale linearly in $\ell_{\rm s}^0$.}
\end{figure}

\section{Longitudinal profiles}\label{sec:long}
To make the content of the similarity hypothesis more palpable, we
report here results for the longitudinal profiles under various wind
and influx conditions obtained by numerical solution of the minimal
model. To simplify the notation we introduce the reference shear
stress $\tau^0$ over the flat bed and corresponding notation for the
saturation length $\ell_{\rm s}^0\equiv \ell_{\rm s}(\tau^0)$, and the
saturated sand flux $q_{\rm s}^0\equiv q_{\rm
s}(\tau^0)$. Figure~\ref{fig:steady}a) combines \emph{steady-state
solutions} (influx = outflux) for various sizes and wind speeds.  The
insets demonstrate exemplarily that all windward profiles are well
represented by

\begin{equation}\label{eq:hx}
  h(x)\approx H_{\rm e} \cos^\alpha \bigl(x/L_\alpha\bigr)\;, 
 \qquad  L_\alpha\equiv L_{\rm e}/\arccos \bigl(2^{-1/\alpha}\bigr)
\end{equation}
with $\alpha\approx 3.0$ for dunes and $\alpha\approx 1.8$ for heaps
($H_{\rm e}$, $L_{\rm e}$ are defined in figure~\ref{fig:sketch}b).
Such collapse of the windward profiles is strongly supported by field
data \cite{sauermann-etal:2000,andreotti-claudin-douady:2002}.
Moreover, the height--length relations for diverse masses and wind
speeds can be superimposed on a single master curve by rescaling
lengths and heights by their values $H_{\rm c}$, $L_{\rm c}$ at the
shape transition. The data points at the lower left indicate that in
units of $\ell_{\rm s}^0$ all heaps have roughly the same length $L_0
\leq L \leq L_{\rm c}\approx 1.5 L_0$, which is of the order of the
minimum unstable wavelength of a flat sand bed
\cite{andreotti-claudin-douady:2002}.  The latter therefore also
defines a minimum size for (steady-state) isolated heaps migrating
over plane bedrock. Figure~\ref{fig:steady}b) demonstrates that both
$L_0$ and $L_{\rm c}$ are linearly related to $\ell_{\rm s}^0$.

\begin{figure}
a)$\!\!\!\!\!\!\!\!$
\includegraphics[width=0.5\columnwidth]{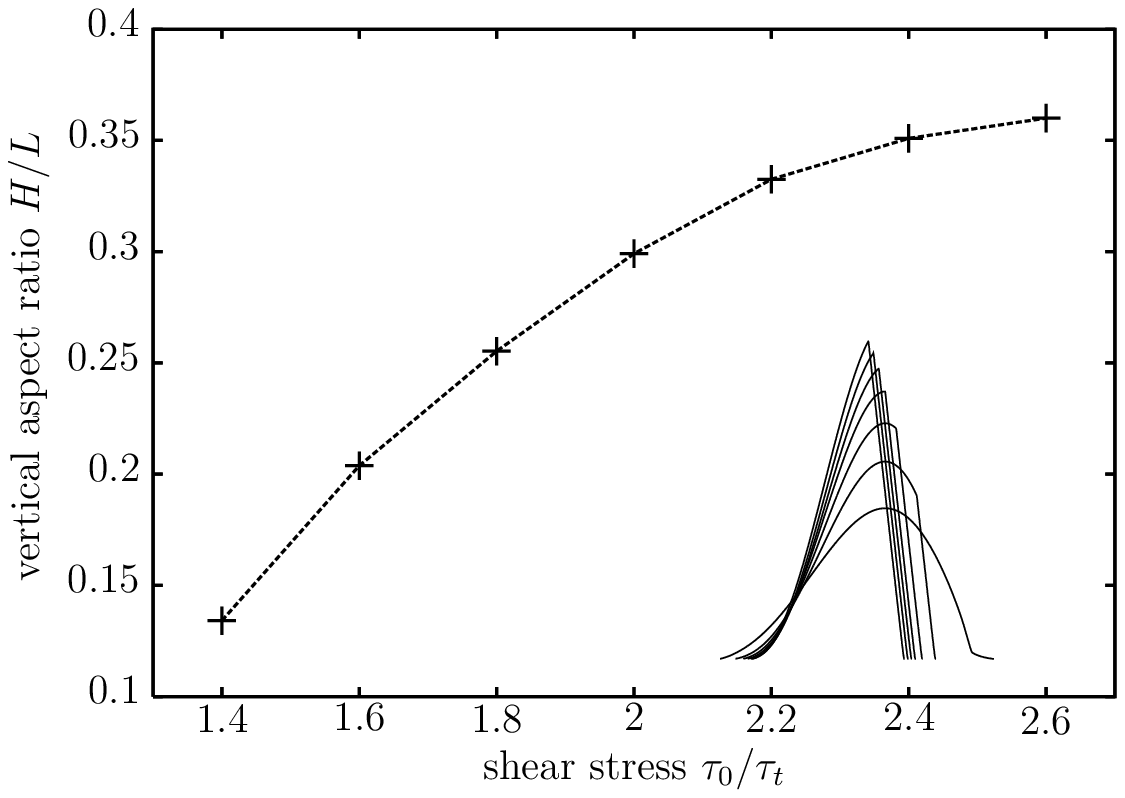}\quad
b)$\!\!\!\!\!\!\!\!$\includegraphics[width=0.5\columnwidth]{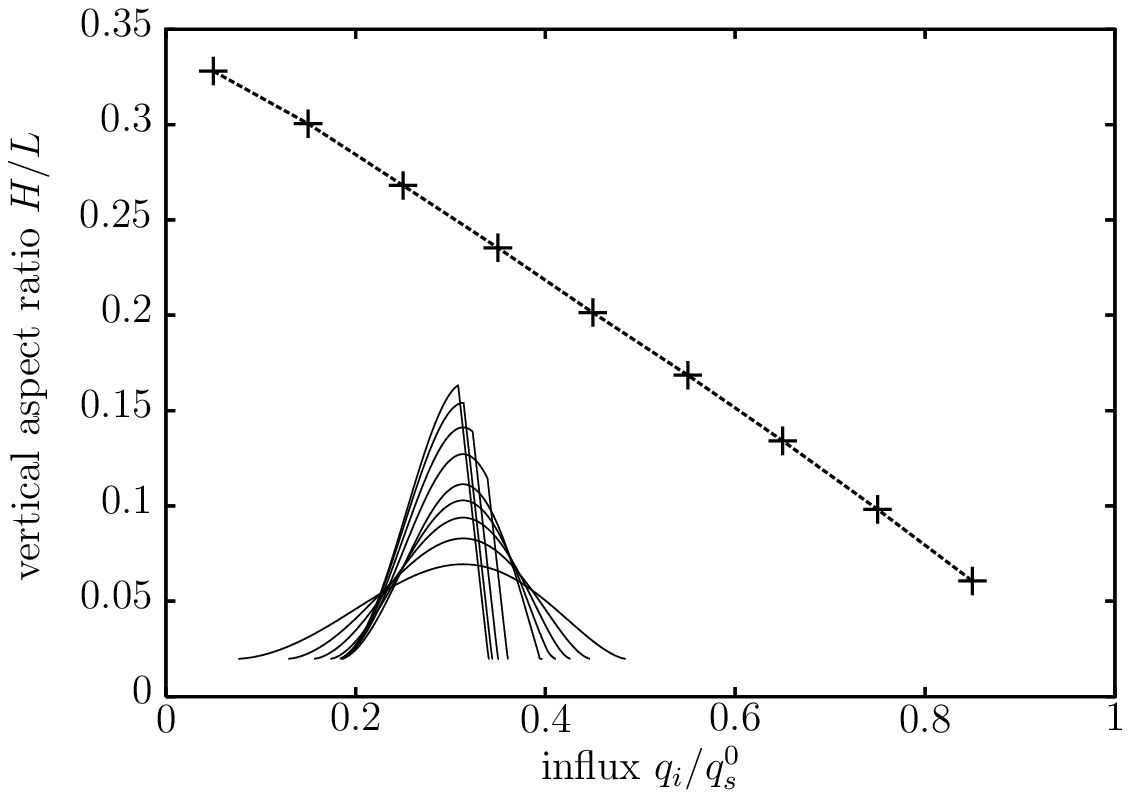}
\caption{\label{fig:unsteady} Longitudinal vertical profiles of given
  mass (225 m$^2$): a) for varying wind shear stress (steady--state
  profiles); b) for given reference shear stress $\tau^0/\tau_{\rm
  t}=2.2$ but varying influx. Again, all windward profiles obey
  equation~(\ref{eq:hx}).}
\end{figure}

The dependence of the critical height $H_{\rm c}$ and critical aspect
ratio $H_{\rm c}/L_{\rm c}$ on the wind strength can be estimated
analytically by observing that the flux over a longitudinal slice has
to vanish upon creation of a slipface. This requires that the minimum
shear stress at the feet of the heap, which deviates from $\tau^0$ by
a factor proportional to $H/L$, equals $\tau_{\rm t}$ if $H=H_{\rm c},
\, L=L_{\rm c}$. Or, $H_{\rm c}/L_{\rm c} \propto 1-\tau_{\rm
t}/\tau^0$, which is in good agreement with our numerical data (not
shown) and mimicked by the aspect ratios of slices of fixed mass
(figure~\ref{fig:unsteady}a). By a similar argument
\cite{kroy-sauermann-herrmann:2002b} one can explain the affine
dependence of the aspect ratio of a slice of given sand mass under a
given wind speed on the influx (figure~\ref{fig:unsteady}b). Note that
contrary to the profiles in figures~\ref{fig:steady}a,
\ref{fig:unsteady}a, the profiles in figure~\ref{fig:unsteady}b are
not in the steady state, but are slowly growing in time.

\section{Transverse profiles}\label{sec:trans}
So far, we have discussed the profiles of longitudinal cuts through
the dune, exclusively.  To understand the transverse dune profile, the
weak lateral coupling of adjacent longitudinal slices obviously has to
be taken into account. In the spirit of a minimal
description\footnote{In the minimal model
\cite{kroy-sauermann-herrmann:2002b}, sand transport is modelled as on
a flat sand bed, the topography enters only via its effect on the wind
\emph{speed}.  Direct slope effects onto the flux (difference between
uphill or downhill hopping, lateral deflection) and onto the local
wind direction (wind deflection) are neglected. Due to the relatively
mild slopes on the windward side, this is a good first
approximation.}\setcounter{fnt}{\value{footnote}}, we regard as the
conceptionally most important contributions lateral grain scattering
in the hopping process and irregular lateral perturbations due to
turbulence. Both effects already exist on a flat bed and give rise to
lateral diffusion \cite{lima-etal:2002}. Though the neglected
systematic slope effects\footnotemark[\value{fnt}] cannot \emph{a
priori} be regarded as small relative to this lateral diffusion, we
prefer to neglect them here to keep the model `minimal' and
analytically tractable. Some investigations of how they affect the
numerical solutions of the minimal model can be found in the
literature
\cite{sauermann:2001,schwaemmle-herrmann:2003b,hersen:2004}.

Within the mentioned restrictions, we make the phenomenologically
plausible assumption that the lateral scattering is roughly
proportional to the longitudinal flux $q_\|$ itself, so that a net
lateral flux will only occcur in response to lateral gradients in
$q_\|$, i.e.\ $q_\perp\approx - \ell_\perp\partial q_\|/\partial y$
\cite{lima-etal:2002}\footnote{Because of the weak dependence of the
  average grain speed on the wind speed as a consequence of the
  non-trivial feedback of the mobilized grains onto the wind
  \cite{sauermann-kroy-herrmann:2001,kroy-sauermann-herrmann:2002b},
  this ansatz is essentially equivalent to a similar approach in
  reference \cite{schwaemmle-herrmann:2003b} involving gradients in
  the density of saltating grains. Compared to the latter it
  represents a considerable simplification, because it does not oblige
  us to introduce separate dynamic equations for the density and
  velocity of the mobilized grains, respectively.}. The new
phenomenological length scale $\ell_\perp$ can be expected to be a
complicated function of the splash process, but will be proportional
to (though supposedly numerically substantially smaller than) the
average saltation length, i.e.\ the effective characteristic hopping
length of the grains, which is predicted to be relatively weakly
sensitive to changes in wind speed
\cite{sauermann-kroy-herrmann:2001,kroy-sauermann-herrmann:2002b} for
the same reason as the grain speed. Therefore, $\ell_\perp$ will be
treated as a constant phenomenological coefficient for the following
considerations. We identify the spatial changes
\begin{equation}\label{eq:e_y}
E_y\equiv \partial q_\perp/\partial y \approx -\ell_\perp\partial^2
q_\|/\partial y^2
\end{equation}
in the lateral flux as a lateral erosion. From this we can deduce the
transverse profile if we can relate the longitudinal flux to the
height profile. Bagnold noticed \cite{Bagnold41} that for a
steady-state dune with migration speed $v$, the ratio $q^{\rm
b}_\|(y)/h^{\rm b}(y)$ of the sand flux over the brink to the height
of the brink must be the same for any longitudinal slice $h_y(x)$ with
slipface, or rather, $q^{\rm b}_\|(y)= \rho_{\rm s}v h^{\rm b}(y)$,
where $\rho_{\rm s}$ is the density of the sand bed.  Given that
practically no sand is lost over the slipface, this immediately
follows from mass conservation. Extending Bagnold's relation to
regions somewhat upwind of the brink, we make the assumption that on
vertical transverse slices $h_x(y)$ ($x$ fixed) the longitudinal flux
can still approximately be parametrized by
\begin{equation}\label{eq:qh}
q_\|(y)\approx \rho_{\rm s}v h_x(y) \;.
\end{equation}
Our conclusions would not change, though, if we allowed for additional
terms $a_x y^2$, $b_x y^4$ with some $x-$dependent coefficients $a_x$,
$b_x$ on the right hand side of this relation and/or terms $\tilde
a_x$, $\tilde b_x y^2$ on the right hand side of
equation~(\ref{eq:e_y}).

We can now derive the profile of transverse slices $h_x(y)$ close to
the crest of a large steady-state barchan dune from the requirement
that the total lateral erosion and deposition (hence the square of
$E_y$) be minimized for a given area of that slice,
\begin{equation}\label{eq:variation}
\delta\int\rmd y\left[ \left(\partial^2 q_\|/\partial y^2\right)^2+\mu
h_x(y)\right]=0\;, \quad h_x(\pm Y)=0\;.
\end{equation}
Here $Y(x)$ denotes the windward base-profile of the dune. Using
equation~(\ref{eq:qh}), the solution for the transverse height profile
is found to be a quartic polynomial that degenerates to a
\emph{parabola}
\begin{equation}\label{eq:hy}
h_x(y)\propto 1-(y/Y)^2 \;,
\end{equation}
in agreement with observations, if we require the slopes $\partial
h_x(y)/\partial y|_Y$ at the edges to be finite
\cite{Hesp98,sauermann-etal:2000}.\footnote{The remark after
equation~(\ref{eq:qh}) gives some hint as to why some of the more
complicated relations between the transverse and longitudinal flux
proposed in the literature
\cite{sauermann:2001,schwaemmle-herrmann:2003b,hersen:2004} do not
give rise to any significant numerical deviations of the transverse
profile from the present result. Relaxing the steady-state assumption
should also be uncritical.}  This boundary condition guarantees that
the otherwise eroded heap-like horns of the dune can be constantly
rebuilt by the lateral flux. Since the total flux leaving the horns of
width $w$ is of the order of $wq_{\rm s}^0$, conservation of horn mass
requires $w(q_{\rm s}^0/2-q_{\rm i})= \int \rmd x \, q_{\perp \rm h}$,
where $q_{\perp \rm h}$ is the $x-$dependent lateral flux \emph{onto}
a horn, i.e., $q_\perp$ taken at $y = \pm (W - w)$. For large dunes,
$w\ll W$ and the integral is estimated as $2\rho_{\rm s} v L\ell_\perp
H/W$. Using $v\propto L^{-1}$
\cite{kroy-sauermann-herrmann:2002b,kroy-guo:2004}, we conclude that
the transverse aspect ratio $H/W$ should, for given wind and influx
conditions, \emph{not depend on dune size} if the horn width $w$ does
not, in accord with field observations
\cite{Hesp98,sauermann-etal:2000}.

\section{Conclusions}
We have presented evidence derived from the minimal model of aeolian
sand dunes for similarity rules relating the profiles of dunes of
different mass and under different environmental conditions.  Many of
the predictions are strongly supported by field measurements and by
recent laboratory experiments.  Some caution is needed, however, when
applying the model to extreme conditions, e.g.\ to dunes on Mars. The
general structure of the (hydrodynamic-type) model equations is
expected to pertain to situations where particular predictions for
their kinetic coefficients cease to apply. Our analysis suggests that
both longitudinal and transverse vertical cuts through a barchan dune
have universal profiles independent of dune mass, wind speed, and sand
supply, because the scale invariance imposed by the turbulent wind is
only weakly broken by the saturation transients in the sand flux. In
general, the aspect ratios of the profiles change with these
parameters, however. For the longitudinal cuts, the dependencies could
be rationalized very well on the basis of the original dynamic
equations of the minimal model, while a minimal, analytically
tractable extension of the model was proposed for the transverse
direction.

\section*{References}


\end{document}